\documentclass[aoas,nameyear,seceqn,dvips]{arximspdf}
\usepackage{graphics}
\doi{10.1214/09-AOAS283}
\volume{4}
\issue{1}
\pubyear{2010}
\firstpage{78}
\lastpage{93}

\begin{document}
\begin{frontmatter}

\title{Optimal experiment design in a filtering context with
application to sampled network data}
\runtitle{Optimal design in filtering context}

\begin{aug}
\author[a]{\fnms{Harsh} \snm{Singhal}\ead[label=e1]{singhal@umich.edu}\corref{}}
\and
\author[a]{\fnms{George} \snm{Michailidis}\thanksref{t1}\ead[label=e2]{gmichail@umich.edu}}
\runauthor{H. Singhal and G. Michailidis}
\affiliation{University of Michigan, Ann Arbor}
\address[a]{Department of Statistics\\
University of Michigan\\
439 West Hall\\
1085 South University\\
Ann Arbor, Michigan 48109\\USA\\
\printead{e1}\\
\phantom{E-mail: }\printead*{e2}}
\thankstext{t1}{Supported in part by NSF
Grants CCR-032557, DMS-05-05535 and DMS-08-06094.}
\pdfsubject{The Annals of Applied Statistics, 2009, Vol.0, No.00, 1-18}
\end{aug}

\received{\smonth{1} \syear{2009}}
\revised{\smonth{8} \syear{2009}}

%
\begin{abstract}
We examine the problem of optimal design in the context of filtering
multiple random walks.
Specifically, we define the steady state E-optimal design criterion and
show that the underlying
optimization problem leads to a second order cone program. The
developed methodology is
applied to tracking network flow volumes using sampled data, where the
design variable
corresponds to controlling the sampling rate. The optimal design is
numerically compared to
a myopic and a naive strategy. Finally, we relate our
work to the general problem of steady state optimal
design for state space models.
\end{abstract}

%

\end{frontmatter}
%
\section{Introduction}
\begin{figure}

\includegraphics{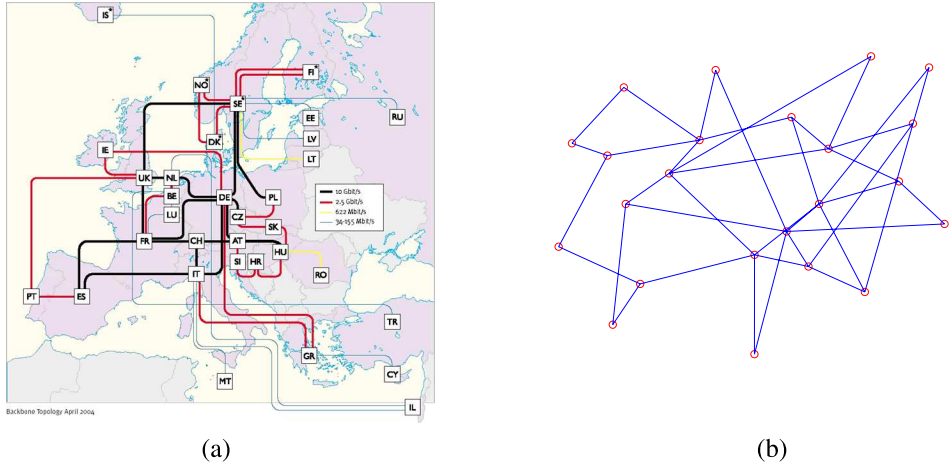}

\caption{Geant Network: \textup{(a)} geographic view (\protect\href{http://www.geant.net}{www.geant.net})
and \textup{(b)} the corresponding logical topology.}
\label{geant}
\end{figure}

\begin{figure}[b]

\includegraphics{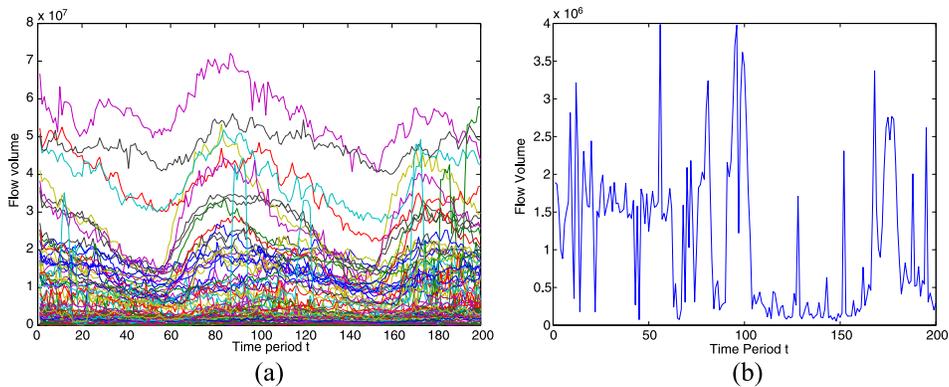}

\caption{Flow volumes: \textup{(a)} all flows and \textup{(b)} one of the lighter flows.}
\label{flowvols}
\end{figure}

Consider a wide area computer network such as the one \mbox{depicted} in
Figure \ref{geant}.
A \textit{flow} is defined as all traffic with common origin and
destination nodes.
Monitoring flow volumes plays an important role in network management
tasks, such
as capacity planning by tracking demands and forecasting traffic,
identifying failures together with their causes and impact,
detecting malicious activity and configuring routing protocols
[\citet{barford02}, \mbox{\citet{soule05}}]. These flow
volumes have been observed to exhibit complicated structure, as seen
in Figure \ref{flowvols}. For example, the highly aggregated flows
usually have diurnal patterns [Figure \ref{flowvols}(a)],
while lighter flows can be extremely noisy [Figure~\ref{flowvols}(b)].
Network traffic is carried on packets that can be observed (and
sampled) at router interfaces, henceforth called \textit{observation
points}. However, during the measurement process, sampling is employed
due to high flow volumes and resource constraints at
routers.

It is increasingly common for such measurement infrastructure to be deployed
in computer networks [\citet{duffieldstatsc}].
Each packet from the aggregate flow at an observation point is sampled
independently with a certain probability
(sampling rate) [\citet{duffield04}]. Typical sampling rates range
between~0.001--0.01. For every packet sampled, its header information is
recorded which allows one
to reconstruct objects of interest, such as volumes of flows with a
particular source and destination
traversing the network. An important issue is how to \textit{select}
(design) the sampling rates across the network
subject to resource constraints, in order to collect the maximum amount
of information on the underlying source-destination flows.
Obviously low sampling rates result in large sampling noise.
One way of achieving lower estimation error with the same sampling
rate is through \textit{filtering}; that is, combining the present measurement
with past measurements to track the time-series of flow volumes.
In designing a sampling scheme for this situation one needs to take
into account measurement noise
and process noise (innovation noise).

While modeling the dynamics of
flow volumes is a challenging task in itself [\citet{park}], we use a simple
random walk model for this purpose. This is a robust enough
model to be useful in a large range of applications and leads to
scalable filters.
We consider the problem of minimizing the (running) estimation
error through optimal design of measurement scheme in the filtering context.
In this paper we take an optimal design of experiment approach to the
above problem
and demonstrate its application to computer network monitoring using
sampled data.

The related research on optimal design has focused on one of the
following scenarios.
There is a large body of work on optimal input design for dynamical systems
[\citet{goodwin}, \citet{titterington}]. There the focus is on parameter
estimation (system identification) rather than filtering, as in this paper.
Another related area is sequential design for nonlinear systems
[\citet{gautier}, \citet{ford}],
where the optimal design depends on values of unknown parameters.
While there are some commonalities, the design problem in a filtering
context is unique
in that the design at any time affects not just the current estimation
error but
also future ones. The problem of optimal sensor placement in control system
literature looks at an equivalent problem [\citet{arbel}, \citet{Chmielewski02}]. However, the
formulation is not in terms
of information matrices and the special case of random walks has not
been analyzed to our knowledge.
More details are provided in Section \ref{sec4}.

The remainder of the paper is organized as follows: in Section \ref{sec2}
we formulate and investigate the idealized problem of optimal
design in the context of filtering for multiple random walks. In
Section \ref{sec3} we study its
application to tracking flow volumes using sampled data. We end with
discussion of
a possible generalization and some comments in Section \ref{sec4}.
\section{Optimal design for multiple random walks}\label{sec2}
Let us first briefly review the concept of E-optimality from classical
design-of-experiment literature
for a simple setting.
Assume we have \textit{independent} observations
%
\begin{equation}
y_i\sim N(x_i,1/m_i),
\label{eqny}
\end{equation}
for $i=1,2,\ldots,n_r$.
The natural estimate for $x_i$ is $\hat{x}_i=y_i$ for all $i$.
It is standard to assume that the inverse variance of observation
noise is
roughly
proportional to design variables.
The inverse variance, $m_i$, can be thought of as the information
collected on parameter $x_i$.
Specifically,
we assume that the
relation between an $n_r\times1$ information vector $m$ and
an $n_o\times1$
vector of design variables $\xi$ is
%
\begin{equation}
m=J\xi.
\label{eqnm}
\end{equation}
For example, suppose there is a library of measurements $z_1,\ldots,z_{n_o}$,
each of which is independently distributed as $z_i\sim N(x_{[i]},\sigma
_i^2 I)$,
where $x_{[i]}$ is a subset of elements of $x$. Let $\xi_i$ be
equal to (or proportional to) the number of independent measurement of
type $i$ (replications of $z_i$)
collected during the experiment. Then, the weighted least squares
estimate $y$ of $x$ can be shown to have distribution given by (\ref{eqny})
and (\ref{eqnm}). The matrix $J$ depends on the the membership
of subsets $x_{[i]}$ and variances $\sigma_i^2$ (assumed known),
for $i=1,\ldots,n_o$.

We assume that the design variables are constrained to be positive and,
in addition, satisfy $n_v$ linear inequality constraints. These can
be written as
$R\xi\leq b$, where $R$ is an $n_v\times n_o$ matrix and $b$ is
$n_v\times1$ vector.
We think of this type of constraint as a budgetary one,
that specifies upper limits on weighted sums of the design variables.

Now the E-optimal design problem is given by
\[
\arg\max_{R\xi\leq b} \min_i m_i.
\]

The objective function, $\min_i m_i$,
is the minimum information over all flows.
Note that this corresponds to minimizing the maximum mean squared
error (MSE)
since $1/m_i$ is the MSE in the estimate of $x_i$. Using maximum
MSE as the objective function corresponds to aiming for the best possible
worst case performance.

\begin{figure}[b]

\includegraphics{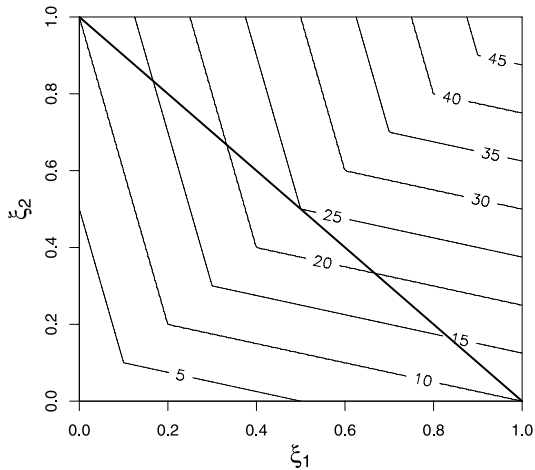}

\caption{Contours of the objective function for E-optimal design.}
\label{objcont1}
\end{figure}

As an example consider the situation where
$m_1=40\xi_1 + 10\xi_2$ and $m_2=10\xi_1 + 40\xi_2$. Further assume the
constraint
\[
\xi_1+\xi_2\leq1.
\]

Figure \ref{objcont1} shows the contours of the objective function,
that is, $\min_i m_i$. The region below the thick line
is the constraint space. As usual, the optimal solution
corresponds to the point where the contour of the objective function is
``tangent''
to the boundary of the constraint space.
It is clear that the optimal design
would be $\xi_1=\xi_2=0.50$, which is also reasonable from the symmetry
of the setup.

We now extend the above criteria to the
optimal design for random walks in a steady state.
Consider a collection of \textit{independent} random walks
\[
x_i(t)=x_i(t-1)+\varepsilon_i(t),
\]
for $i=1,\ldots,n_r$ and $t=1,2,\ldots.$
The term $\varepsilon_i(t)$ is called the innovation noise and
we assume that $\operatorname{Var}(\varepsilon_i(t))=\sigma_i^2$, which
is referred to as the innovation variance.
Further, suppose we have noisy observations
\[
y_i(t)=x_i(t)+\eta_i(t).
\]
Let
$\operatorname{Var}(\eta_i(t))=1/m_i$.
As before, we assume the
relation between observed information and design variables
to be $m=J\xi$, with $n_r\times n_o$ matrix $J$ assumed known.

The estimates of interest
in this case are the ones obtained through filtering
\[
\hat{x}_i(t)=E[x_i(t)|y_i(t),y_i(t-1),\ldots].
\]
Let $s_i(t)=\operatorname{Var}(\hat{x}_i(t)|y_i(t),y_i(t-1),\ldots)$. Further,
let $\tilde{m}_i=\lim_{t\to\infty}1/s_i(t)$ when it exists.
We will refer to this as the \textit{steady state information}.
When the innovation and measurement noise, $\varepsilon_i(t)$ and $\eta_i(t)$
respectively, are Gaussian, the optimal filter corresponds to a Kalman filter
and in this case the steady state always exists [\citet{harvey}].
For the remainder of the paper we will assume that
$\varepsilon_i(t)$ and $\eta_i(t)$
are independent mean 0 Gaussian random variables.
If $s_i(t|t-1)=\operatorname{Var}(\hat{x}_i(t)|y_i(t-1),y_i(t-2),\ldots)$, then
the Kalman filter update equations give us
%
\begin{equation}
s_i(t|t-1)=s_i(t-1)+\sigma_i^2
\label{eqnkf1}
\end{equation}
and
%
\begin{eqnarray}
s_i(t)^{-1}&=&s_i(t|t-1)^{-1}+m_i\\
&=&\biggl(\frac{1}{s_i(t-1)^{-1}}+\sigma^2_i\biggr)^{-1}+m_i.
\label{eqnkf2}
\end{eqnarray}
Note that, given $s_i(0)$, $\sigma_i^2$ and $m_i$, one can calculate $s_i(t)$
at any time $t$ by iterating the above equations. Further, the choice of
$m_i$ impacts $s_i(t)$ not just for a specific~$t$ but for all $t$.
Thus,
\[
\tilde{m}_i=\biggl(\frac{1+\sigma_i^2\tilde{m}_i}{\tilde{m}_i}\biggr)^{-1}+m_i
\]
or
\[
\sigma_i^2\tilde{m}_i^2-\sigma_i^2m_i\tilde{m}_i-m_i=0.
\]

Hence,
\[
\tilde{m}_i=\frac{m_i\sigma_i^2+\sqrt{m_i^2\sigma_i^4+4m_i\sigma_i^2}}{2
\sigma_i^2}.
\]

We define the
steady state E-optimal design problem as
\[
\arg\max_{R\xi\leq b}\min_i \tilde{m}_i.
\]

\begin{figure}[t]

\includegraphics{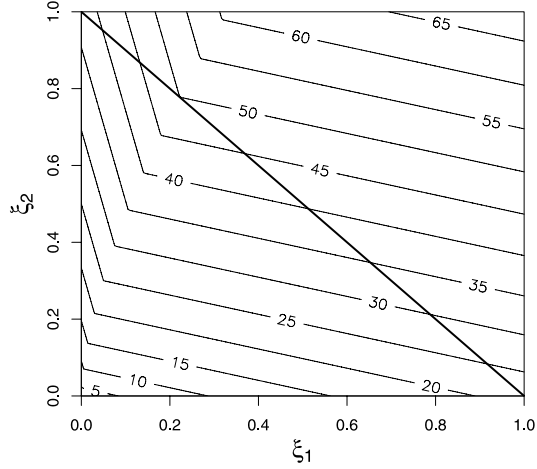}

\caption{Contours of the objective function for steady state E-optimal design.}
\label{objcont2}
\end{figure}

As an example consider the same setting as above, with
$m_1=40\xi_1 + 10\xi_2$ and $m_2=10\xi_1 + 40\xi_2$.
Further, let the innovation noise
be characterized by
$\sigma_1=0.1$ and $\sigma_2=0.2$. As before, we assume the design
constraint
\[
\xi_1+\xi_2\leq1.
\]
Notice in Figure \ref{objcont2} that even though there is symmetry in the
measured information, the first random walk is smoother than the second one
and, hence, less measurement resources need to be allocated to it.
\subsection{Optimization for the steady state E-optimal design}
We establish next the main technical result of the paper, that the
steady state E-optimal design
problem is a second order cone program.
First, we introduce a new variable $\theta$ as the lower bound
for the steady state information over all flows.
To solve the steady state E-optimal design problem,
we have to maximize $\theta$ subject to
%
\begin{equation}
\frac{m_i\sigma_i^2+\sqrt{m_i^2\sigma_i^4+4m_i\sigma_i^2}}{2
\sigma_i^2}\geq\theta,
\label{origcons}
\end{equation}
for $i=1,\ldots,n_r$ and
\[
R\xi\leq b.
\]
Equation (\ref{origcons}) can be equivalently written as
%
\begin{equation}
\theta^2\leq m_i\biggl(\theta+\frac{1}{\sigma_i^2}\biggr),
\label{hyperbolic}
\end{equation}
which is a hyperbolic constraint [\citet{boydsocp}].
Thus, this problem can be cast as a second order cone program (see the
\hyperref[app]{Appendix}
for a review of second order cone programs and the representation
of the above optimization in canonical form). Such optimization
programs can be solved efficiently through interior point methods
[\citet{opt}],
software implementations of which are commonly available
[\citet{software1}, \citet{software2}].

\subsection{Myopic approach}
In the following, we present a greedy alternative to the steady state
optimal design.
As before, assume $y_i(t)=x_i(t)+\eta_i(t)$.
Further, we assume that
$\operatorname{Var}(\eta_i(t))=1/m_i(t)$;
that is, we allow for time varying design variables $\xi(t)$
with $m(t)=J\xi(t)$.
As before, $s_i(t)=\operatorname{Var}(\hat{x}_i(t)|y_i(t),y_i(t-1),\ldots)$.
Define the information at time $t$ to be given by $\tilde{m}_i(t)=1/s_i(t)$.
Note that $\tilde{m}_i(t)$ is a function of $\xi(t),\xi(t-1),\ldots.$

The myopic E-optimal design at time $t$ is defined as
\[
\arg\max_{R\xi(t)\leq b}\min_i \tilde{m}_i(t).
\]

Note that the objective function only involves $\tilde{m}_i(t)$,
that is, the information at time $t$. However, the choice of $\xi(t)$ impacts
not just $\tilde{m}_i(t)$ but also $\tilde{m}_i(t+1),\tilde
{m}_i(t+2),\ldots,$
due to the iterative nature of Kalman filtering. Since it ignores this ``long
term impact,'' we refer to this scheme as myopic.
Equation~(\ref{eqnkf2}) implies that
\[
\tilde{m}_i(t)=s_i(t|t-1)^{-1}+J\xi(t).
\]
As before, a new variable $\theta$ can be
introduced to lower bound $\tilde{m}_i(t)$
which gives a new set of constraints
\[
s_i(t|t-1)^{-1}+J\xi(t)\geq\theta,
\]
in addition to the original constraint $R\xi(t)\leq b$. Now the
objective is
to maximize~$\theta$ with the optimization variables being $\theta$ and
$\xi(t)$.
Since both the objective
function and the constraints are linear in $\xi(t)$ and $\theta$,
the above optimization is a linear program.
Not surprisingly, the myopic optimal design is a much easier
problem than the steady-state optimal design even in more general
settings as noted in Section~\ref{sec4}.
Note that since the sampling rates are allowed
to vary with time, the myopic optimal design
may have an objective function larger than
the steady state optimal case. However, as the objective of
optimization is to maximize
present information with no regard to impact on future information,
such a scheme can not be guaranteed to perform well in the long run.

\section{Application to tracking flow volumes}\label{sec3}
The ideas developed above can be used for designing the sampling rate
in a computer network for tracking flow volumes.
As mentioned in the introduction, we will use the random walk model for
flow volumes due to its simplicity and robustness.
%
\begin{figure}

\includegraphics{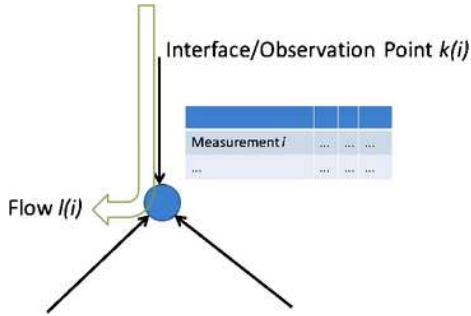}

\caption{Schematic representation of an observation, observation points
and flows.}
\label{schema}
\end{figure}

Suppose there are $n_r$ origin-destination flows in a network.
Let $x_i(t)$ be the volume of the $i$th flow in the $t$-the time
interval, for
$i=1,\ldots,n_r$. These flow volumes are tracked using
sampled data which are noisy. Recall that flows are sampled at router
interfaces, which we refer to as observation points.
In the past, a systematic sampling scheme was the dominant technology,
but truly random sampling technologies have recently become available
and are
commonly deployed [\citet{duffield04}].
All flows traversing an {observation point} (router
interface) experience the same sampling rate.
Each incoming edge at a node in Figure~\ref{geant}(b)
is an interface of the corresponding router.
Each router typically has multiple interfaces and
each flow may traverse multiple observation points due to multi-hop
paths and multi-path routing.

Suppose there are $n_o$ observation points
on the network where sampled data on flows can be collected.
Further, assume that
sampling rates of $\xi=
(\xi_1,\ldots,\xi_{n_o})'$ are used at observation points $1,\ldots,n_o$,
respectively. Any given observation point $k\in\{1,\ldots,n_o\}$ generates
estimates for $g_k$ elements of $x(t)$, that is, the number
of flows that go through that node. Thus,
a total of $n_g=\sum_{k=1}^{n_o}g_k$
measurements are available in each time interval $t$,
say, $z_1(t),\ldots,z_{n_g}(t)$, which need to be optimally combined
to get the
required estimates.
Consider the router in Figure \ref{schema}.
Assume that $k(i)$ is the observation point
at which the $i$th measurement is collected, $i=1,\ldots,n_g$,
and $l(i)$ the corresponding
flow. Thus, $k(\cdot)\dvtx\{1,\ldots,n_g\}\rightarrow\{1,\ldots,n_o\}$ and
$l(\cdot)\dvtx\{1,\ldots,n_g\}\rightarrow\{1,\ldots,n_r\}$.
Further, let
\[
E\bigl[z_i(t)|x_{l(i)}(t)\bigr]=x_{l(i)}(t)
\]
and for the moment assume
%
\begin{equation}
\operatorname{Cov}\bigl(z_i(t)|x_{l(i)}(t)\bigr)=\mu_{l(i)}/\xi_{k(i)},
\label{approx}
\end{equation}
where $\mu_i=E[x_{i}(t)]$.
The exact sampling mechanism and approximation involved in the
above relation are described in Section \ref{sec3.2}.
Thus, in vector notation we get
%
\begin{equation}
E[z(t)|x(t)]=Lx(t),
\label{v1}
\end{equation}
where $L$ is a $n_g\times n_r$ matrix with $L_{ij}=1$ only if $l(i)=j$
(i.e., $i$th measurement corresponds to $j$ flow)
and $0$ otherwise and
%
\begin{equation}
\operatorname{Cov}(z(t)|x(t))=D,
\label{v2}
\end{equation}
where $D$ is a $n_g\times n_g$ diagonal matrix, with
$[D]_{ii}=\mu_{l(i)}/\xi_{k(i)}$.
Using (\ref{approx}), the inverse of $D$ is given by
$D^{-1}=\sum_k\xi_k\Psi_k$, where $\Psi_k$, $k=1,\ldots,n_o$, are
$n_g\times n_g$
diagonal matrices with their $i$th element given by
%
\begin{equation}\label{lin}
[\Psi_k]_{ii}=\cases{
1/\mu_{l(i)},&\quad
$\mbox{if } k=k(i),$\cr
0,&\quad $\mbox{otherwise}.$}
\end{equation}

Let $y(t)$ be the general least squares estimate of $x(t)$, under
equation (\ref{v1}) and~(\ref{v2}).
Thus,
%
\begin{eqnarray}
\operatorname{Cov}(y(t)|x(t))&=&(L'D^{-1}L)^{-1}\label{mscov}\\
&=&\biggl(\sum_k(L'\Psi_kL)\xi_k\biggr)^{-1}.
\end{eqnarray}

From the definition of~$L$, it follows that the $j$th elements
of any two columns of~$L$ cannot be nonzero simultaneously. Thus,
the matrix in (\ref{mscov}) is diagonal. Further,
\[
\operatorname{Diag}(\operatorname{Cov}(y(t)|x(t)))^{-1}=m=J\xi,
\]
where
\[
[J]_{ik}=L_{\cdot,i}'\Psi_kL_{\cdot,i} .
\]
We will refer to the above as the linear model.

Sampling is employed in network flow measurements
because measurement resources like CPU time and available storage
are limited. Typically, all observation points (router interfaces)
belonging to a particular router share these resources.
We assume that the sampling rates are constrained to lie in a convex
polygon $R\xi_t\leq b$.
This includes the case where
the sum of sampling rates on the interfaces of
a router is bounded above by the budget for that router.
We will focus on this constraint for the rest of the paper.
In this case, the constraints are given as one linear inequality for
each router.

For the available data, we set up the performance evaluation as follows.
We use the Geant network topology, which has $n_v=23$ nodes (routers) and
$37\times2$ bidirectional edges.
The available data [\citet{uhlig}]
correspond to flow volumes over time. Each time
interval is equal to 15 minutes. The original data set spans 4 months,
but we focus on the first 200 time intervals to avoid severe
non-stationarities inherent in an evolving network.
Further, we focus on the top 25\% of measured
flows by volumes since one is typically interested in tracking heavy flows.
This corresponds to $n_r=76$ flows. We assume that sampled
data can be collected at each incoming edge of a router and, thus,
we have $n_o=37\times2$ observation points. We assume that these
flows are routed through minimum distance paths, which is a common
routing mechanism in
wide area networks [\mbox{\citet{peterson}}]. This leads to $n_g=163$
and the routing information gives us the
mapping $l(\cdot)$ and hence
the matrix $L$. Matrix $L$ is $163 \times76$, matrices $D,\Psi_1,\ldots,
\Psi_{74}$ are all $163 \times163$. We assume that the sum
of sampling rates on all interfaces of a router is bounded above by 0.01,
that is, $b_i=0.01$ and $R_{ij}=1$ if observation point
$j$ is an interface of router $i$ and 0 otherwise. Thus, matrix $R$
is $23 \times74$. Finally, we estimate the $\sigma_i^2$
and $\mu_i$ parameters associated with the flow volume processes, and
assume they are available for filtering purposes and measurement design.
As we have argued, both the steady state optimal and myopic design problems
are standard optimization programs and once they are written as such, any
standard optimization package [\citet{software1}, \mbox{\citet{software2}}]
can be used to solve them numerically.

For the purpose of comparison, we also define a naive sampling scheme
as follows.
For any given router, an equal sampling rate is allocated to
every interface that carries any of the 76 flows of interest.
This allocation is done so as to
make the corresponding budget constraint tight.
For example, suppose the $i$th
router has~5 interfaces, but only 4 of them are traversed
by one of the 76 flows of interest. In this case, each of the latter 4
interfaces
will be allocated a sampling rate of $b_i/4$, while the remaining interface
will be allocated a sampling rate of 0.

\subsection{Performance of various sampling schemes for the linear model}
\begin{figure}[b]

\includegraphics{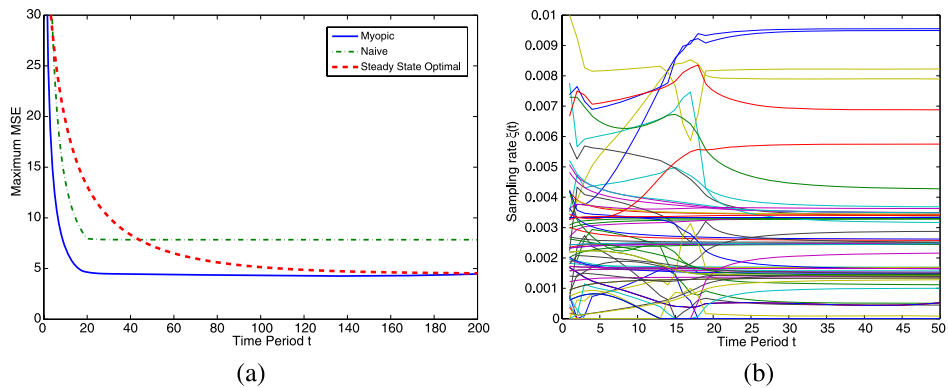}

\caption{Performance of various sampling schemes [panel \textup{(a)}] and
sampling rates at various
interfaces under a myopic scheme [panel \textup{(b)}].}
\label{linperf}
\end{figure}

\begin{figure}

\includegraphics{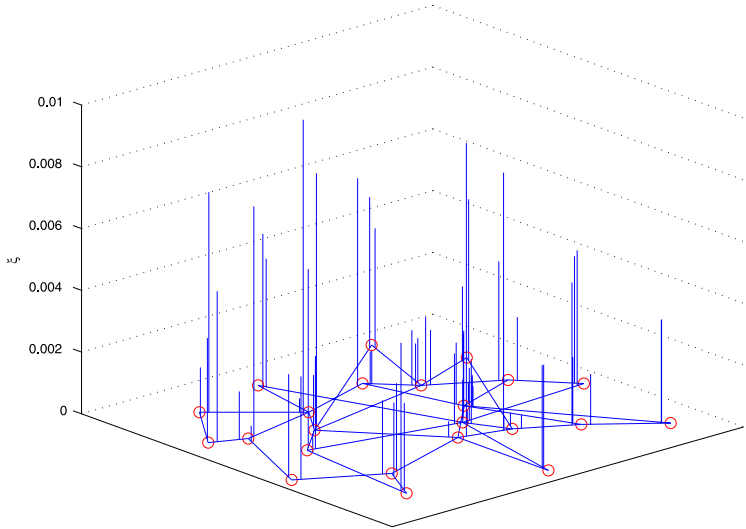}

\caption{Spatial view of steady state optimal sampling rates.}
\label{spatial}
\end{figure}

Figure~\ref{linperf}(a) shows the value of the maximum MSE as a
function of time.
Note that as information accumulates over
time, we obtain an improvement in performance under all three sampling
mechanisms,
myopic, naive and steady state optimal.
Here performance is measured as the maximum of $s_i(t)$ over all flows,
calculated using equations (\ref{eqnkf1}) and (\ref{eqnkf2}).
Surprisingly, both the myopic and steady state optimal sampling
mechanisms perform equally well in the steady state and
achieve a 42\% improvement over
the naive sampling in the steady state. Figure \ref{linperf}(b) shows
that the
myopic optimal sampling rates at all observation points reach a steady state.
Figure \ref{spatial} shows the value of steady state sampling rates
at various router interfaces in the network topology.
Even though the myopic scheme has the flexibility of time varying
sampling rates, if the sampling rates do reach a steady state, its performance
can clearly be no better than the steady-state optimal scheme.
However, as Figure \ref{linperf} shows, in this case,
the additional flexibility permits
the myopic scheme to reach steady state performance faster than the
steady state
optimal one.

\subsection{Departures from the linear model: performance with geant
data}\label{sec3.2}
A more detailed model for flow volumes and sampled measurements would
have to include
significant departures from the linear model assumed above.
First, the true flow volumes clearly have more structure than independent
random walks, as seen in Figure \ref{flowvols}.
In applying the above ideas to the Geant data, we will
investigate their robustness to the independent random walk assumption.

A more serious departure is the following.
Suppose that a flow with volume $X$ in a certain
time interval is sampled at a rate $\xi$.
If the number of sampled packets
is $N$, then the usual (approximate maximum likelihood) estimate of
flow volume is $Z\equiv N/\xi$.
The variance of measurement noise can be shown to be
$\operatorname{Var}(Z|X)\simeq X/\xi$ [\citet{duffield}]. Thus, $\mu_i$ in
(\ref{approx}) is actually equal to the unknown $x_i(t)$.

\begin{figure}

\includegraphics{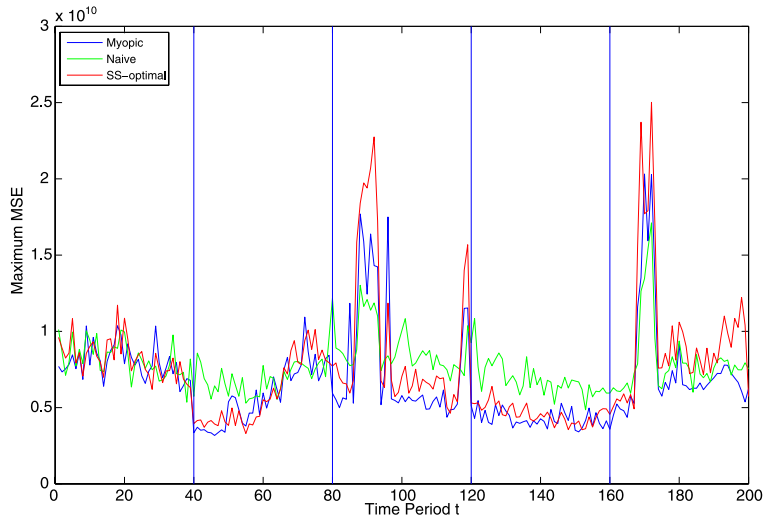}

\caption{Performance of various sampling schemes using
batch sequential design with flow volumes from the Geant data.}
\label{batch}
\end{figure}

The observation above implies that in applying the presented techniques
to sampled network data,
one would have to rely on an approximate model for measurements $z_i(t)$.
We will follow an approach similar to batch sequential design [\citet{gautier}].
Assume that the sampling rates are to be held constant for
a batch of contiguous time intervals.
At the beginning of each batch, we use the most recent estimate $\hat{x}_i(t-1)$
in place of $\mu_i$ in (\ref{lin}) for sampling rate design. For
filtering purposes,
we employ a Kalman filter with $\hat{x}_i(t-1)$ in place of $\mu_i$ in
equation~(\ref{lin}) at each time $t$.
We replace the budget constraint inequalities $R\xi\leq b$ with the
corresponding
equalities $R\xi=b$ to force full utilization of available resources.
For routers that are traversed by at least one of the 76 flows of interest,
we introduce additional equality constraints as follows. Design variable
$\xi_k$ for an interface~$k$
not traversed by one of the 76 flows of interest is constrained to
be identically 0.
Figure \ref{batch}
shows the performance of different sampling schemes averaged over
200 realizations of sampled data. The sampled data emulate the exact
sampling mechanism described above (with respective sampling
rates) with the Geant data treated as the underlying
(unobserved) flow volumes.

Sampling rates were adjusted only at the beginning of a 40 time period block
and were held constant over each block. In the first block, the sampling
rates were forced to be the same as the naive scheme irrespective of
the sampling mechanism under study. Notice that for low values
of the objective function (maximum mean squared error) the myopic and steady
state allocations perform better than the naive allocation. On the
other hand, when
the maximum mean squared error spikes, the naive allocation performs better,
indicating robustness to model departures. The median (over time
periods 41 to 200)
of maximum MSE for myopic, naive and steady-state optimal sampling
is $5.46\times10^9$, $7.49\times10^9$ and $6.14\times10^9$, respectively.
Thus, the myopic scheme performs better than the steady state optimal scheme,
which in turn performs better than the naive scheme.

\begin{figure}

\includegraphics{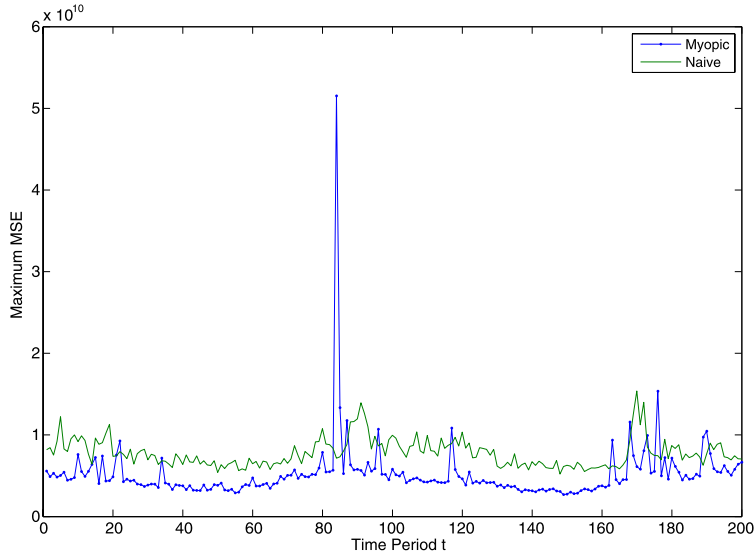}

\caption{Performance of fully time varying myopic and naive sampling mechanism
with flow volumes from Geant Data.}
\label{eachp}
\end{figure}
Finally, we look at the performance of the myopic allocation when
the above scheme is employed with a block size of just one time
interval; that is,
sampling rates were adjusted at the beginning of each time period using the
myopic scheme. The results
are displayed in Figure \ref{eachp}. As before, the current estimate
of flow volumes is used
in place of $\mu_i$ in equation (\ref{lin})
for both filtering and
myopic sampling scheme design. The myopic sampling scheme can be seen
to perform
better than the naive version in most time periods. The median
(over time periods 1 to 200) of maximum MSE is $4.50\times10^9$
and $7.44\times10^9$ for myopic and naive sampling, respectively.

\section{Discussion and future work}\label{sec4}
The \textit{specification} of
the steady state optimal design problem can be easily generalized to
linear dynamical systems.
Such systems are described by a pair of equations [\citet{harvey}].
The
state transition equation can be written as
\[
x(t)=Cx(t-1)+w(t),
\]
where $\textrm{Cov}(w(t))=W$.
The observation equation can be written as
\[
y(t)=
Lx(t)+ \varepsilon(t).
\label{obs}
\]
Assume $\textrm{Cov}(\varepsilon(t))=\Psi(\xi)^{-1}$,
where $\Psi(\cdot)$ is a linear function and $\xi$ is the
value of design variables.

For the above dynamical system a Kalman filter can be used to
iteratively compute $E[x(t)|y(t),y(t-1),\ldots]$.
Let the steady state estimation error covariance
be $\Sigma=\tilde{M}^{-1}$ (assuming the system is
observable [\citet{harvey}]).
Then, $\tilde{M}$ satisfies the Algebraic Riccati equation:
%
\begin{equation}
\tilde{M}=(C\tilde{M}^{-1}C'+W)^{-1}+L'\Psi(\xi)L .
\label{eqnare}
\end{equation}
Such equations have no analytic solution in general.

The steady state optimal design problem can now be defined as
\[
\arg\max_{R\xi\leq b}f(\tilde{M}) ,
\]
where $f(\cdot)$ is an appropriate scalarization of the information matrix
[\citet{fedorov}].
An interesting
open problem is to solve the above
optimization efficiently in the absence of an analytic solution to (\ref
{eqnare}).
The sensor placement problem in control system literature [\citet
{arbel}] is equivalent, though
not identical. The Newton-type algorithm proposed in [\citet{arbel}]
for this problem requires the solution
of the Algebraic Riccati equation at each iteration of the algorithm.
It would be desirable to develop more
efficient algorithms.

In summary, we have shown that steady state E-optimal design for
random walks is a second order cone program.
We have illustrated numerically that the performance of the Kalman
filter can be significantly improved by incorporating an optimal
experimental design.
The linear state space
model is of general interest and one would like
to investigate the steady state optimal design problem described above.
Finally, from a practical
point of view, it would be useful to extend these ideas to nonlinear filtering.

\begin{appendix}\label{app}
\section*{Appendix: Optimization review}
In this section we summarize the concepts of
second order cone programs and hyperbolic constraints
from \citet{boydsocp}. We also present
the steady state optimal design problem in the canonical form.

A second order cone program is defined as
\begin{eqnarray*}
&&\mbox{minimize } f'x
\\
&&\mbox{subject to}\qquad\Vert P_ix+q_i\Vert\leq r_i'x+s_i, \qquad i=1,\ldots,N.
\end{eqnarray*}
Here, $x\in\mathbb{R}^n$ is the optimization variable, and
the problem parameters are $f\in\mathbb{R}^n$, $P_i\in\mathbb
{R}^{n_i\times
n}$, $q_i\in\mathbb{R}^{n_i}$, $r_i\in\mathbb{R}^n$ and $s_i\in\mathbb{R}$.
The norm in the constraints is the standard Euclidean norm.
A second order cone program is a standard convex program and algorithms
to numerically solve it are well studied and implemented in computational
software.

A constraint of the form
\[
w^2\leq xy , \qquad x\geq0,\qquad y\geq0
\]
is called hyperbolic. The above can be shown to be equivalent to
\[
\left\|\pmatrix{
2w\cr
x-y}
\right\|
\leq x+y.
\]

Using the above representations, we can write the steady state optimal
design problem
as a canonical second order cone program as follows. Equation
(\ref{hyperbolic})
can be equivalently written as
\[
\left\|\pmatrix{
\theta\vspace*{2pt}\cr
m_i-\theta-\dfrac{1}{\sigma_i^2}
}
\right\|
\leq m_i+\theta+\frac{1}{\sigma_i^2}.
\]

Thus, the steady state optimal problem is a second order cone program
with $N=n_r+n_v$, $x'=(\theta,\xi_1,\ldots,\xi_{n_o})$,
$f'=(-1,0,\ldots,0)$,
\[
P_i=\pmatrix{
2&0&\cdots&0\cr-1&J_{i,1}&\cdots& J_{i,n_o}},
\]
for $i=1,\ldots,n_r$ and $P_i=\mathbf{0}$, for $i=n_r+1,\ldots,n_r+n_v$,
$q_i'=(0,-1/\sigma_i^2)$, for $i=1,\ldots,n_r$ and
$q_i=\mathbf{0}$, for $i=n_r+1,\ldots,n_r+n_v$, $r_i'=(1, J_{i,1},
\ldots, J_{i,n_o})$,
for $i=1,\ldots,n_r$ and $r_i'=(0,-R_{i-n_r,1},\ldots,-R_{i-n_r,n_o})$,
for $i=n_r+1,\ldots,n_r+n_v$,
$s_i=1/\sigma_i^2$,
for $i=1,\ldots,n_r$ and $s_i=b_{i-n_r}$,
for $i=n_r+1,\ldots,n_r+n_v$.
\end{appendix}

\section*{Acknowledgments} The authors would like to thank
the Editor Steve Fienberg, the Associate Editor
and two referees for helpful comments and suggestions.

\printaddresses

\end{document}